# Comment: Bayesian Checking of the Second Level of Hierarchical Models: Cross-Validated Posterior Predictive Checks Using Discrepancy Measures

Michael D. Larsen and Lu Lu

## 1. INTRODUCTION

We compliment Bayarri and Castellanos (BC) on producing an interesting and insightful paper on model checking applied to the second level of hierarchical models. Distributions of test statistics (functions of the observed data not involving parameters) for judging appropriateness of hierarchical models typically involve nuisance (i.e., unknown) parameters. BC (2007) focus on ways to remove the dependency on nuisance parameters so that test statistics can be used to assess models, either through $p$-values or Berger's relative predictive surprise (RPS). They demonstrate shortcomings in terms of very low power of posterior predictive checks and a posterior empirical Bayesian method. They also demonstrate better performance of their partial posterior predictive ($ppp$) method over a prior empirical Bayesian method. Methods of Dey et al. (1998), O'Hagan (2003) and Marshall and Spiegelhalter (2003) also are compared.

Methods are contrasted in terms of whether they require proper prior distributions, how many measures of surprise (one per group or one total) are produced, and the degree to which data are used twice in estimation and testing. Their preferred method ($ppp$) can use improper prior distributions, which are referred to as objective, produces a single measure of surprise for each test statistic, and avoids double use of the data. For the models and statistics considered, in comparison to the alternatives presented, $ppp$ has a more uniform null distribution of $p$-values and more power versus alternatives.

In this discussion, we suggest that cross-validated posterior predictive checks using discrepancy measures hold some promise for evaluating complex models. We apply them to O'Hagan's data example, provide some comments on the paper and discuss possible future work.

## 2. CROSS-VALIDATED POSTERIOR PREDICTIVE CHECKS USING DISCREPANCY MEASURES

Suppose there are data for $I$ groups: $X_i, i = 1, \ldots, I$, where $X_i = (X_{ij}, j = 1, \ldots, n_i)$. The unknown parameters in the first level in group $i$ are $\theta_i$: $f(X_i|\theta_i)$ independently. The parameters in the second level of the model are $\eta$: $\pi(\theta|\eta) = \prod_{i=1}^{I} \pi(\theta_i|\eta)$. The prior distribution on $\eta$ is $\pi(\eta)$. Let $D(X, \theta, \eta)$ be a generalized discrepancy measure. If $D(X, \theta, \eta) = D(X)$, then it is a test statistic. Examples are given in the next section for the normal-normal model considered by BC (2007). Cross-validated posterior predictive model checking using a discrepancy measure is implemented as follows. Separately for each $i = 1, \ldots, I$:

1. Generate $M$ values ($m = 1, \ldots, M$) from the posterior distribution of $\eta|X_{(-i)}$; call them $\eta_{(-i)}^m$, where $X_{(-i)}$ represents all the data without group $i$. Generating values of $\eta$ will be accomplished in many cases through iterative simulation methods that will generate values of $\theta_{(-i)}$, where $\theta_{(-i)}$ is the collection of group parameters excluding group $i$: $f(\eta|X_{(-i)}) = \int f(\eta, \theta_{(-i)}|X_{(-i)}) \, d\theta_{(-i)} \propto \int \pi(\eta) \pi(\theta_{(-i)}|\eta) f(X_{(-i)}|\theta_{(-i)}) \, d\theta_{(-i)}$.
2. Generate values $\theta_i^m$ of $\theta_i$ given the hyperparameters $\eta_{(-i)}^m$ independently from $\pi(\theta_i|\eta_{(-i)}^m)$, $m = 1, \ldots, M$.

M. D. Larsen is Associate Professor, L. Lu is graduate student, Department of Statistics and Center for Survey Statistics & Methodology, Iowa State University, Snedecor Hall, Ames, Iowa 50011, USA e-mail: larsen@iastate.edu; icyemma@iastate.edu.

TABLE 1
*Posterior predictive p-values for individual groups and the whole population*

| Discrepancy | Group 1 | Group 2 | Group 3 | Group 4 | Group 5 | Whole population |
|---|---|---|---|---|---|---|
| Overall $X^2$ | 0.568 | 0.857 | 0.261 | 0.747 | 0.287 | 0.483 |
| 1st Level $X^2$ | 0.547 | 0.893 | 0.140 | 0.893 | 0.202 | 0.496 |
| 2nd Level $X^2$ | 0.512 | 0.594 | 0.567 | 0.518 | 0.403 | 0.513 |
| $\text{Max}_{j\in\{1,\ldots,n_i\}} X_{ij}$ | 0.476 | 0.851 | 0.060 | 0.847 | 0.143 | — |
| $\text{Max}_{j\in\{1,\ldots,n_i\}} |X_{ij} - \theta_i|$ | 0.610 | 0.839 | 0.113 | 0.923 | 0.283 | — |
| $\text{Max}_{j\in\{1,\ldots,n_i\}} |X_{ij} - \mu|$ | 0.682 | 0.820 | 0.286 | 0.897 | 0.151 | — |
| $\text{Max}_i |\bar{X}_i - \mu|$ | — | — | — | — | — | 0.493 |

3. Generate replicate data $X_i^m$ independently from $f(X_i|\theta_i^m)$, $m = 1, \ldots, M$.
4. Compute the proportion of times out of $M$ that $D(X_i^m, \theta_i^m, \eta_{(-i)}^m)$ is greater than $D(x_i, \theta_i^m, \eta_{(-i)}^m)$, $m = 1, \ldots, M$.

This proposal allows the use of objective prior distributions, is relatively easy to implement in many hierarchical models, avoids double use of data in group $i$ for evaluating the model for group $i$, and allows many test statistics and discrepancy measures to be used based on one set of simulations of $\eta$ and $\theta$. On the negative side, this procedure may lose some power for some statistics compared with *ppp*, but likely much less so than regular posterior predictive checks. The use of more flexibly defined discrepancies, however, could produce relatively powerful evaluations for some aspects of some models. The proposal requires more computing than regular posterior predictive checks and faces issues of multiplicity in testing. The method is applied in Section 3 and followed by discussion in Section 4.

## 3. O'HAGAN'S EXAMPLE

O'Hagan's data [see Section 5 of BC (2007)] are used to study the performance of model checking based on regular and cross-validated posterior predictive checks utilizing various discrepancy measures. The model being fit is a two-level normal-normal hierarchical model. Notation is the same as in BC (2007).

Different discrepancy measures relate to various parts of the model. The overall $X^2$ discrepancy, defined by $\sum_{j=1}^{n_i} \frac{(X_{ij} - \mu)^2}{(\sigma^2 + \tau^2)}$ for group $i$, measures the adequacy of two levels as a whole. The first and second level $X^2$ discrepancies, defined as $\sum_{j=1}^{n_i} \frac{(X_{ij} - \theta_i)^2}{\sigma^2}$ and $\frac{(\theta_i - \mu)^2}{\tau^2}$ for group $i$, detect the inadequacy of the first- and second-level models, respectively. The three measures above also can be summed across groups, $i = 1, \ldots, I$. The maximum absolute deviation of a group average from the overall center is $\text{Max}_i |\bar{X}_i - \mu|$ and quantifies fit of the whole model. The maximum value $\text{Max}_{j\in\{1,\ldots,n_i\}} X_{ij}$ and the minimum value $\text{Min}_{j\in\{1,\ldots,n_i\}} X_{ij}$ in group $i$ are sensitive to extremes within groups. The maximum absolute deviations of observations from the group mean in group $i$, $\text{Max}_{j\in\{1,\ldots,n_i\}} |X_{ij} - \theta_i|$, relates to spread about the mean within group $i$. The maximum absolute deviation of observations from the overall mean in group $i$, $\text{Max}_{j\in\{1,\ldots,n_i\}} |X_{ij} - \mu|$, relates to adequacy of both levels in the model.

For the regular posterior predictive checks noninformative prior distributions for parameters $\sigma^2$, $\mu$ and $\tau^2$ were used: $\pi(\mu) \propto 1$, $\pi(\sigma^2) \propto 1/\sigma^2$ and $\pi(\tau^2) \propto 1/\tau$ (or equivalently $\pi(\tau) \propto 1$). Table 1 shows the posterior predictive $p$-values for individual groups and the whole population. As observed by BC (2007), suffering from the double use of data, none of the discrepancy measures detect any evidence of incompatibility between the observed data and the null model for individual groups or for the population as a whole.

Table 2 shows the $p$-values based on cross-validated posterior predictive checks for individual groups. The model fits the data from groups 1, 2 and 4 very well. For group 3, the $p$-values based on the first-level $X^2$ discrepancy is 0.016, which indicates slight inadequacy of the first-level model. This is not surprising due to the extreme observation 4.10. The impact of this unusual observation in group 3, given a model of equal spread in each group, also is detected by the discrepancy measure $\text{Max}_{j\in\{1,\ldots,n_i\}} |X_{ij} - \theta_i|$, which has a $p$-value of 0.023. Despite the concern about the first-level model in group 3, discrepancy measures focused on the second level and the model overall do not detect any problem. This is consistent with the fact that the mean and spread in group 3 are not extreme compared with the other groups.

TABLE 2
*Cross-validated posterior predictive p-values for individual groups*

| Discrepancy | Group 1 | Group 2 | Group 3 | Group 4 | Group 5 |
|---|---|---|---|---|---|
| Overall $X^2$ | 0.653 | 0.804 | 0.520 | 0.730 | **0.007** |
| 1st Level $X^2$ | 0.168 | 0.315 | **0.016** | 0.291 | **0.000** |
| 2nd Level $X^2$ | 0.577 | 0.656 | 0.654 | 0.585 | **0.007** |
| $\text{Max}_{j \in \{1,\ldots,n_i\}} X_{ij}$ | 0.641 | 0.723 | 0.373 | 0.759 | **0.005** |
| $\text{Max}_{j \in \{1,\ldots,n_i\}} |X_{ij} - \theta_i|$ | 0.203 | 0.333 | **0.023** | 0.411 | **0.002** |
| $\text{Max}_{j \in \{1,\ldots,n_i\}} |X_{ij} - \mu|$ | 0.715 | 0.819 | 0.472 | 0.841 | **0.006** |

For group 5, all discrepancies detect the inadequacy of the hierarchical model. This makes sense since group 5 has a very extreme group mean of 4.44, which is almost three times the other group means, and has at least one relatively extreme observation of 6.32, which is almost twice the overall within-group standard deviation away from the group mean. Note that even if *p*-values for group 5 were multiplied by 5 or 6 to deal with multiplicity of testing, the result would still be less than 0.05 for all the various discrepancies.

Now we consider improving the proposed hierarchical model by using more robust distributions for modeling the outlying group and extreme observations. Since we have seen slight inadequacy in the first-level model for groups 3 and 5 and serious inadequacy in the second-level model for group 5, we might consider using Student-t distributions to accommodate the unusual observations and the extreme group mean parameter in the hierarchical model.

To perform a robust analysis, we replace the normal distributions by Student-t distributions with fixed degrees of freedom $\nu_1 = 3$ and $\nu_2 = 2.2$ in the first and second levels of the hierarchical model. The cross-validated posterior predictive *p*-values assuming Student-t distributions in both levels of model are shown in Table 3. The two-level robust Student-t model successfully accommodates the unusual observation in group 3 and almost accommodates the extreme observation in group 5. But it does not fully address the inadequacy of the second-level model for fitting group 5's data. Given this result, one might suggest treating group 5 as being generated from a normal distribution with a shifted location parameter or an inflated variance parameter. One could also consider using another model, such as one of BC's (2007) alternative models in their Section 3.6. If there were more groups with higher means, then fitting a mixture of normal distributions in the second level might be an option.

Degrees of freedom greater than 2 are used because such *t*-distributions have finite variances. A little bit of experimenting was done to choose the degrees of freedom. Larger degrees of freedom had less success (slightly) of fitting the data, but made little difference in posterior distributions of parameters or in results in Table 3. If the degrees of freedom are thought of as parameters, then posterior variance will be quite high with this few groups.

TABLE 3
*Cross-validated posterior predictive p-values for individual groups assuming Student-t distributions for both levels in the hierarchical model*

| Discrepancy | Group 1 | Group 2 | Group 3 | Group 4 | Group 5 |
|---|---|---|---|---|---|
| Overall $X^2$ | 0.680 | 0.856 | 0.493 | 0.822 | 0.074 |
| 1st Level $X^2$ | 0.211 | 0.376 | 0.081 | 0.381 | 0.060 |
| 2nd Level $X^2$ | 0.636 | 0.676 | 0.667 | 0.639 | **0.022** |
| $\text{Max}_{j \in \{1,\ldots,n_i\}} X_{ij}$ | 0.581 | 0.664 | 0.320 | 0.734 | 0.070 |
| $\text{Max}_{j \in \{1,\ldots,n_i\}} |X_{ij} - \theta_i|$ | 0.295 | 0.450 | 0.117 | 0.501 | 0.122 |
| $\text{Max}_{j \in \{1,\ldots,n_i\}} |X_{ij} - \mu|$ | 0.732 | 0.877 | 0.440 | 0.891 | 0.134 |

## 4. SOME COMMENTS ON THE PAPER AND DISCUSSION

From the above analysis we can see that it is useful to employ various discrepancies to measure the overall performance and the specific assumptions of the model. Cross-validated posterior predictive checking allows the use of many discrepancies focused on various aspects of the model and avoids the double use of data. It is also useful for assessing individual small groups or areas that are inconsistent with the model. Extensions to multilevel models, models with covariates and generalized linear models should be possible. See Gelman (2004) and Gelman et al. (2005) and references therein for other examples of model diagnostics that use flexibility in defining evaluations to advantage.

The framework of test statistics only for checking models is less flexible and requires more effort; test statistics of BC's (2007) Section 3.3 required some refinement of procedures in Appendix C. The authors should be commended on their efforts and explanations; their results show a definite advantage over the other methods in their article in these applications.

The authors state that they intend the model checks to be preliminary in order to avoid model elaboration and (possibly) averaging. It seems unlikely to us that there would not be value in using such methods for further study of models past an initial stage. Indeed, it might be the case that unusual patterns might be detectable only after models reach a certain level of complexity. We agree with the authors that assessing total uncertainty through an elaborate model selection and refinement procedure is a challenge that deserves more study.

An issue for future work with model assessment is multiplicities: the use of multiple test statistics or discrepancy measures to evaluate a single model and tests concerning individual groups. Multiplicity in testing will affect power and distribution of $p$-values. One could recommend selecting one discrepancy to assess each part of a model and avoid too much overlap and redundancy. We agree with BC (2007) that in cases with many discrepancy measures and, in particular, many groups, simple Bonferroni corrections might decrease power too much; in such cases investigation of methods from statistical genetics (small $n$, large $p$) might be helpful. As a side note, it would not be particularly hard to simulate $p$-value distributions and power for cross-validated posterior predictive $p$-values under the scenarios of BC (2007) with or without adjustment for multiplicity.

In order to implement cross-validated posterior predictive checking one must sample the posterior distribution while leaving out groups of data. When the number of groups or areas is large, the computation needed for reanalyzing the model without each group or area could be time consuming. To avoid refitting the model without each group, methods such as importance weighting and importance resampling could be used to approximate the posterior distribution that would be obtained if the analysis were repeated with leaving out the group. See Stern and Cressie (2000), Marshall and Spiegelhalter (2003) and references therein in this regard.

Again we wish to thank authors for a stimulating paper that demonstrates a method that seems quite effective and clearly states issues involved.

## ACKNOWLEDGMENTS

The authors wish to thank the editor Ed George for the opportunity to discuss this article and Bayarri and Castellanos for helpful comments on the discussion. This work was supported in part by Iowa's State Board of Education and a dissertation award from the American Education Research Association.